\documentstyle[twoside,fleqn,espcrc2]{article}
\input{epsf}

\def\ffrac#1#2{{\textstyle\frac{#1}{#2}}}
\def\la{\mathrel{\mathpalette\fun <}}
\def\ga{\mathrel{\mathpalette\fun >}}
\def\fun#1#2{\lower3.6pt\vbox{\baselineskip0pt\lineskip.9pt
\ialign{$\mathsurround=0pt#1\hfill##\hfil$\crcr#2\crcr\sim\crcr}}}

\def\APM {{\rm APM}}
\def\CDM {{\rm CDM}}

\def\en {{\rm end}}

\def\lin {{\rm lin}}

\def\mpc {h^{-1} {\rm Mpc}}

\def\impc {h {\rm Mpc}^{-1}}
\def\rmB {{\rm B}}
\def\rmc {{\rm c}}
\def\rmd {{\rm d}}
\def\rme {{\rm e}}
\def\rmm {{\rm m}}
\def\sta {{\rm start}}

\newcommand{\AmS}{{\protect\the\textfont2
  A\kern-.1667em\lower.5ex\hbox{M}\kern-.125emS}}

\title{Evidence for an inflationary phase transition from the
LSS and CMB anisotropy data}

\author{J.Barriga\address{\small{Institut d'Estudis Espacials de Catalunya, 
     IEEC, Institut de Ci\`encies de l'Espai/CSIC, \\
     Edf. Nexus-201 - c/ Gran Capit\`a 2-4, 08034 Barcelona, Spain.}}%
     E. Gazta\~naga$^{a,}$\address{\small{
     INAOE, Astrof\'{\i}sica,\\
     Tonantzintla, Apdo. Postal 216 y 51, Puebla 7200, Mexico.}},
     M.G. Santos$^{c}$ and S. Sarkar\address{\small{Theoretical Physics, 
     University 
     of Oxford, \\
     1 Keble Road, OX1 3NP, UK}}}

\begin{document}

\begin{abstract}
In the light of the recent Boomerang and Maxima observations of the CMB
which show an anomalously low second acoustic peak, we reexamine the
prediction by Adams et al (1997) that this would be the consequence of a
'step' in the primordial spectrum induced by a spontaneous symmetry
breaking phase transition during primordial inflation. We demonstrate that
a deviation from scale-invariance around $k\sim0.1h$~Mpc$^{-1}$ can
simultaneously explain both the feature identified earlier in the APM
galaxy power spectrum as well the recent CMB anisotropy data, with a
baryon density consistent with the BBN value. Such a break also allows a
good fit to the data on cluster abundances even for a critical density
matter-dominated universe with zero cosmological constant.
\end{abstract}
\maketitle

\section{Introduction}

It is commonly assumed by astronomers that inflation predicts a
scale-invariant `Harrison-Zeldovich' (H-Z) spectrum of scalar density
perturbations, $P_0(k)\propto k^n$ with $n=1$. Adams et al. (1997)
realized that the spectrum may not
be even {\em scale-free} because the rapid cooling of the universe
during primordial inflation can result in spontaneous symmetry
breaking phase transitions which may interrupt the slow roll of the
inflaton field for brief periods. This is in fact inevitable in
models based on $N=1$ supergravity, the phenomenologically successful
extension of the Standard Model of particle physics and the effective
field theory below the Planck scale. During inflation, the large
vacuum energy breaks (global) supersymmetry giving otherwise massless
fields (`flat directions') a mass of order the Hubble parameter
\cite{dfn84,cfkrs84}
, causing them to evolve rapidly to the asymmetric
global minima of their scalar potential. Such `intermediate scale'
fields are generic in models derived from superstring/M-theory and
have gauge and Yukawa couplings to the thermal plasma so are initially
confined at the symmetric maxima of their scalar
potentials. Consequently it takes a (calculable) finite amount of
cooling before the thermal barrier disappears and they are free to
evolve to their minima 
\cite{y86}
. When a symmetry breaking
transition occurs, the mass of the inflaton field changes suddenly
(through couplings in the K\"ahler potential), temporarily violating
the slow-roll conditions and interrupting inflation (Adams et
al. 1997).\footnote{When (re)heating occurs at the end of inflation
such fields may again be forced back to the symmetric maximum,
undergoing symmetry breaking a second time when the universe cools
down to the electroweak scale in the radiation-dominated era and
driving a late phase of `thermal inflation' 
\cite{ls96}
.} Thus the
density perturbation is expected to have a (near) H-Z spectrum for the
first $\sim10$ e-folds of expansion followed by one or more sudden
departures from scale-invariance lasting $\sim1$ e-fold. In order for
such spectral features to be observable in the LSS or CMB, it is of
course necessary that they occur within the last $\sim50$ e-folds of
inflation, corresponding to spatial scales going up to the present
Hubble radius $H_0^{-1}\sim3000\mpc$. Since the density perturbation
can be observed on scales from the Hubble radius down to $\sim1\mpc$,
corresponding to about 8 e-folds of expansion, it would be not
unreasonable to expect at least one such spectral break to be seen
today.

\section{Reconstructing the primordial spectrum}

\subsection{Primordial spectrum from the APM}

\begin{figure}[htb]
\vspace{9pt}
\centerline{\epsfxsize=8truecm\epsfbox{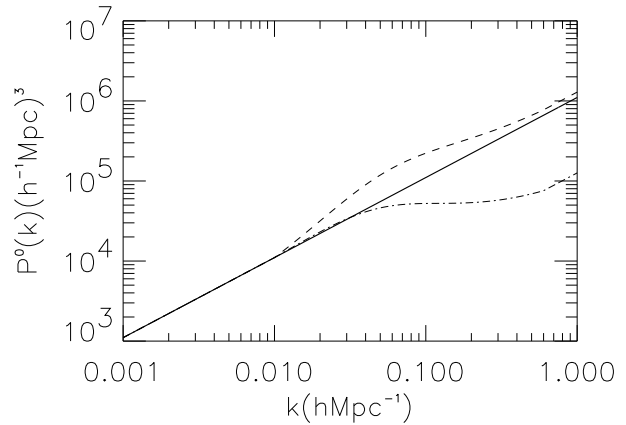}}
\caption[junk]{\small{Reconstruction of the primordial density power spectrum
 from the APM data, adopting a CDM shape parameter $\Gamma$ of 0.5
 (dot-dashed line) and 0.2 (dashed line). A Harrison-Zeldovich
 spectrum (full line) is shown for comparison.}}
\label{fig1}
\end{figure}

There are several arguments (reviewed in the Appendix of \cite{bgss00}) that 
on scales $0.01\mpc \la k \la 0.6\impc$, which are at most weakly non-linear,
the APM galaxy power spectrum $P_\APM(k)$ 
\cite{be93}
 is an {\em
unbiased} (or moderately linearly biased) tracer of the mass. 
The linear power spectrum
recovered under this assumption from $P_\APM(k)$ is well fitted in this range 
by \cite{bg96}:$P_{\lin}(k) 
\simeq \frac{7\times10^{5} k\,(\mpc)^3}{\left[1+(k/k_\rmc)^2\right]^{1.6}}$
where $k_\rmc=150(H_0/c)\simeq0.05\,\impc$. Here we will also
use the common convention: $P_{\lin}(k)\equiv P^0(k) ~ T^2(k)$ where $P^0(k)$ 
is the primordial spectrum of matter fluctuations and
$A$ is the (dimensionful) normalisation constant. 
From the expressions above we estimate 
$P^0(k)$ to be :
\begin{eqnarray}
 P^0(k) = \left\{ \begin{array}{ll}
  A_1 k, & \mbox{$k<k_1$,}\\
  A k~ \frac{1/T_{\CDM}^2(k)}
   {\left[1+(k/k_\rmc)^2\right]^{1.6}}, & \mbox{$k_1 \leq k \leq k_2$,}\\
  A_2 k, & \mbox{$k_2<k$,}
\end{array} \right.
\label{pkprim}
\end{eqnarray}
 $T_{\CDM}=[1+{ak+(bk)^{3/2}+(ck)^2}^{\nu}]^{-1/\nu}$ given by \cite{be84}, 
where $a=6.4\Gamma^{-1}\mpc$, $b=3\Gamma^{-1}\mpc$, $c=1.7\Gamma^{-1}\mpc$, 
$\nu=1.13$ and the `shape parameter' 
$\Gamma =\Omega_\rmm{h}\rme^{-[\Omega_\rmB(1+\sqrt{2h}/\Omega_\rmm)-0.06]} $. 
$A=7\times10^{5} (\mpc)^3$ 
and $A_1$ and $A_2$ are such that make $P^0(k)$ continuous. For
 the cosmological parameters which  define $\Gamma $ we consider the 
observationally indicated values $h\sim0.5-0.8 $ and $\Omega_\rmB = 
(0.019^{+0.0013}_{-0.0012})h^{-2}$ \cite{flsv98} (for
further discussion see \cite{bgss00})).
Figure~\ref{fig1} shows the recovered primordial spectrum (\ref{pkprim})
for two choices of $\Gamma$ corresponding to the sCDM model ($\Gamma =0.5$) 
 and a low density variant. Note that the $\Gamma=0.5$ reconstruction has
significantly less power than a scale-free H-Z spectrum on scales
$k\ga 0.1\impc$, while the $\Gamma=0.2$ reconstruction is closer to a
H-Z spectrum but has relatively more power. Thus the latter possibility does 
{\em not} give a good fit to the Boomerang/MAXIMA data with the value (given 
above) of the baryon density from big bang nucleosynthesis (the reader is 
referred to \cite{bgss00} for further comments on this).

\subsection{Fits to the CMB and LSS data}

\begin{figure}[t]
\vspace{9pt}
\centerline{\epsfxsize=8truecm\epsfbox{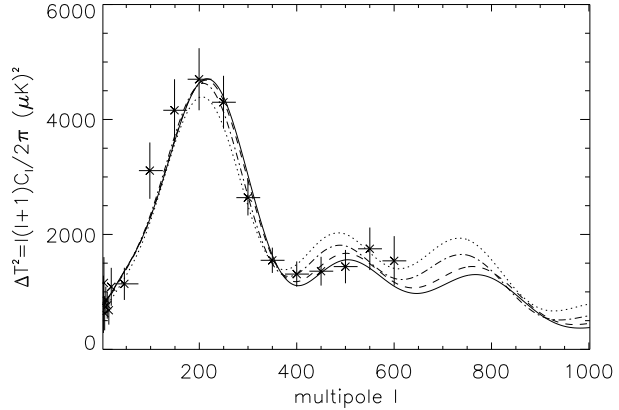}}
\caption[junk]{\small{CMB angular power spectra for the best-fit models in
 Table~\ref{tab4} with $\Omega_{\Lambda}=0$ (continuous line),
 $\Omega_{\Lambda}=0.2$ (dashed line), $\Omega_{\Lambda}=0.4$
 (dot-dashed line), and $\Omega_{\Lambda}=0.6$ (dotted line). All models
 have the BBN baryon density $\Omega_{\rmB}=0.019/h^2$. The data shown
 are from COBE and Boomerang.}}
\label{fig5}
\end{figure}

We parameterise the ``step'' in the primordial power spectrum (see
Figure~\ref{fig1}) as:
\begin{displaymath}
 P^0(k)= \left\{ \begin{array}{ll}
         A k,          & k \leq k_\sta \\
         C k^{\alpha}, & k_\en \leq k \leq k_\sta \\
         B k,          & k \geq k_\en \\
                \end{array}\right.
\label{pkmodel}
\end{displaymath}
where $k_\sta$ and $k_\en$ mark the begining and end of the break from
H-Z spectra with amplitudes $A$ and $B$. (The values of $C$ and
$\alpha$ are specified by the other parameters.) In the multiple
inflation model 
\cite{ars97b}
, the actual form of the spectrum during
the phase transition is difficult to calculate since the usual
`slow-roll' conditions are violated. However a robust expectation is
that $\ln(k_\en/k_\sta)\sim1$ because the field undergoing the
symmetry-breaking phase transition evolves exponentially fast to its
minimum. The ratio of the amplitudes $A/B$ is determined by the
(unknown) superpotential couplings of the field undergoing the phase
transition but is expected to exceed unity (i.e. there is a decrease
in the power). Table~\ref{tab4} shows the result of imposing the BBN
constraint and the Hubble parameter range mentioned in the subsection before 
 and also requiring that $0.5\leq\ln(k_\en/k_\sta)\leq2$. We
see that the data now prefer lower values of $\Omega_\Lambda$ (and
$h$).In particular the value
$\Omega_\Lambda\sim0.7$ favoured by the SN~Ia data 
\cite{scp99,hzs98} is {\em not} permitted.
\begin{table}
\begin{tabular}{|l|l|l|l|l|l|l|} \hline
$\Omega_\Lambda$ & $h$ & $k_{start}$ & $A/B$ & 
$\sigma_8$ & $\chi^2$ \\ \hline
\hline
0.0  &  0.50  &  0.07   &  4.3-6.5  &  0.65-0.73  &
  11.9\\ \hline
0.2  &  0.55  &  0.06   &  3.6-5.6  &  0.69-0.78  &
  10.0\\ \hline
0.3  &  0.60  &  0.06   &  3.4-5.0  &  0.76-0.85  &
 10.3\\ \hline
0.4  &  0.65  &  0.05   &  3.1-4.6  &  0.77-0.87  &
 13.1\\ \hline
0.5  &  0.70  &  0.05   &  2.6-4.1  &  0.81-0.93  &
 18.8\\ \hline
0.6  &  0.75  &  0.05   &  2.0-2.8  &  0.90-0.99  &
 29.7\\ \hline
\end{tabular}
\caption[junk]{\small{Parameters for best fits to CMB+APM data
 with the BBN constraint on the baryon density.}}
\label{tab4}
\end{table}
Figure~\ref{fig5} shows the fit to the CMB
data for these models.

\subsection{Implications for cluster abundances}

\begin{figure}[t]
\vspace{9pt}
\centerline{\epsfxsize=8truecm\epsfbox{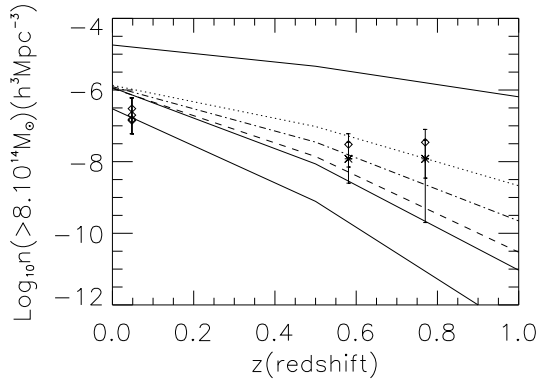}}
\caption[junk]{\small{Number density of massive clusters with
$M> 8 \times 10^{14}$ solar masses as a function of redshift. 
Points with errorbars correspond to the
observations as depicted in Bahcall \& Fan 1998. The thick continuous lines
correspond to the P-S prediction for the flat H-Z model with
$\Omega_\Lambda=0.2$ (ie $\Omega_m \simeq 0.8$), normalised to COBE (top line)
and  scaled to match the cluster abundance at $z=0$ (bottom line). 
The middle lines correspond to the best
fitted models in Table \ref{tab4} with  $\Omega_\Lambda=0$ (continuous line),
 $\Omega_\Lambda=0.2$ (dashed line), $\Omega_\Lambda=0.4$ (dot-dashed
 line), and $\Omega_\Lambda=0.6$ (dotted line).}}
\label{fig8}
\end{figure}

Table~\ref{tab4} shows the value of the variance $\sigma(R)$ in
(dark matter) fluctuations (normalised to the CMB), over a sphere of
size $R=8\mpc$: $\sigma^2 (R)=1/(2\pi^2)\int_0^\infty W^2(kR)P_{lin}(k)~T^2(k)~
 k^2 \rmd k $ using a `top hat' smoothing function,
$W(kR)=3[\ffrac{\sin(kR)}{(kR)^3}-\ffrac{\cos(kR)}{(kR)^2}]$. The tabulated
values are smaller than the values we would obtain when using a H-Z primordial
spectrum, since the $\Gamma $ values for these cosmological models are quite 
high which, according to the APM data (see Figure 1), implies a decrease power
 for the relevant scales. In Figure \ref{fig8} we show the Press-Schechter 
 predictions for the redshift evolution of cluster number densities compared to
the observations, as presented in Bahcall \& Fan (1998). Here we are
displaying the predictions for clusters with $M> 8 \times 10^{14}$ solar
masses. The flat H-Z $\Omega_\Lambda=0.2$ (ie $\Omega_m \simeq 0.8$) model 
normalisedd to 
COBE (top thick continuous line) fails to match the observations of cluster
abundances, a fact that has been used to rule out this model. The figure shows
 how the models fitted in Table
\ref{tab4} do match the redshift evolution, specially after allowing for the
possibility mentioned above of having a slightly smaller $\sigma_8$ values to
better match the $z=0$ data. Note that these predictions have not been adjusted
in any way to predict cluster abundances, all parameters were fixed by the
APM-CMB fit. 

\section{Discussion}

Allowing the primordial spectrum not to be 
scale-invariant has interesting consequences. In our analysis we find
that high values of $\Omega_m$ in a flat universe (and therefore low values of
$\Omega_\Lambda$) are compatible with CMB and LSS data. The standard
interpretation of LSS data (eg APM $P(k)$ and cluster abundances) favours of a
low $\Gamma \sim \Omega_m h \sim 0.2$ for a H-Z spectrum.  But these
observations can also be explained with larger values of $\Omega_m$ if we
allow for a break in the primordial spectrum.  Our best fit values in
Table~\ref{tab4} imposing the BBN constraint mentioned above and the Hubble
parameter constraint in range, prefer higher values of $\Omega_m$ (and lower
values of $\Omega_\Lambda$).  The required spectral break decreases with
increasing $\Omega_\Lambda$ but one cannot do without such a break.  The value
$\Omega_\Lambda\sim0.7$ favoured by the SN~Ia data 
\cite{scp99,hzs98} is {\em not} permitted, while it is possible to have a 
universe with no cosmological constant and  $\Omega_m \simeq 1$.

The hypothesis of a primordial density perturbation with broken
scale-invariance is eminently falsifiable. The
ongoing 2DF and SDSS redshift surveys can confirm or rule out such a
feature in the power spectrum of galaxy clustering, while the
forthcoming MAP mission will determine whether {\em all} the secondary
acoustic peaks in the CMB angular spectrum are indeed suppressed as
expected. Broken scale-invariance has a natural explanation in a phase
transition occuring during inflation as expected in supersymmetric
theories. If established this would provide the first direct
connection between astronomical data and physics at very high
energies.(The reader is referred to \cite{bgss00} for further
details, references and acknowledgements).

\section*{Acknowledgments}
J.B. would like to thank INAOE for their warm hospitality. J.B. and E.G.
acknowledge support by grants from IEEC/CSIC and DGES(MEC)(Spain) project
PB96-0925, and Acci\'on Especial ESP1998-1803-E. The work of M.G.S. was 
supported by the Funda\c c\~ao para a Ciencia e a Tecnologia under program
PRAXIS XXI/BD/18305/98.

\end{document}